\RequirePackage{lineno}
\documentclass[aps,prl,twocolumn,superscriptaddress]{revtex4}

\usepackage{mathrsfs,amssymb,graphics,subfigure,threeparttable}
\usepackage{grap hicx}
\usepackage{amssymb}
\usepackage{enumerate}
\usepackage{amsmath}
\usepackage{fancyhdr}
\usepackage{bm}
\usepackage[squaren]{SIunits}
\usepackage{xcolor}
\usepackage{soul}

\begin{document}


\title{Ultra-compact attosecond X-ray free-electron lasers utilizing unique beams from plasma-based acceleration and an optical undulator}


\author{Xinlu Xu}
\email[]{xuxinlu@pku.edu.cn}
\affiliation{State Key Laboratory of Nuclear Physics and Technology, and Key Laboratory of HEDP of the Ministry of Education, CAPT, Peking University, Beijing, 100871, China}
\affiliation{Beijing Laser Acceleration Innovation Center, Beijing 100871, China}
\author{Jiaxin Liu}
\affiliation{State Key Laboratory of Nuclear Physics and Technology, and Key Laboratory of HEDP of the Ministry of Education, CAPT, Peking University, Beijing, 100871, China}
\author{Thamine Dalichaouch}
\affiliation{Department of Physics and Astronomy, University of California Los Angeles, Los Angeles, CA 90095, USA}
\affiliation{Department of Electrical Engineering, University of California Los Angeles, Los Angeles, CA 90095, USA}
\author{Frank S. Tsung}
\affiliation{Department of Physics and Astronomy, University of California Los Angeles, Los Angeles, CA 90095, USA}
\author{Zhen Zhang}
\affiliation{SLAC National Accelerator Laboratory, Menlo Park, CA 94025}
\author{Zhirong Huang}
\affiliation{SLAC National Accelerator Laboratory, Menlo Park, CA 94025}
\author{Mark J. Hogan}
\affiliation{SLAC National Accelerator Laboratory, Menlo Park, CA 94025}
\author{Xueqing Yan}
\affiliation{State Key Laboratory of Nuclear Physics and Technology, and Key Laboratory of HEDP of the Ministry of Education, CAPT, Peking University, Beijing, 100871, China}
\affiliation{Beijing Laser Acceleration Innovation Center, Beijing 100871, China}
\affiliation{Collaborative Innovation Center of Extreme Optics, Shanxi University, Shanxi 030006, China}
\author{Chan Joshi}
\affiliation{Department of Electrical Engineering, University of California Los Angeles, Los Angeles, CA 90095, USA}
\author{Warren B. Mori}
\affiliation{Department of Physics and Astronomy, University of California Los Angeles, Los Angeles, CA 90095, USA}
\affiliation{Department of Electrical Engineering, University of California Los Angeles, Los Angeles, CA 90095, USA}

\date{\today}

\begin{abstract}
Accelerator-based X-ray free-electron lasers (XFELs) are the latest addition to the revolutionary tools of discovery for the 21st century. The two major components of an XFEL are an accelerator-produced electron beam and a magnetic undulator which tend to be kilometer-scale long and expensive. Here, we present an ultra-compact scheme to produce 10s of attosecond X-ray pulses with several GW peak power utilizing a novel aspect of the FEL instability using a highly chirped, pre-bunched and ultra-bright electron beam from a plasma-based accelerator interacting with an optical undulator. The self-selection of electrons from the combination of a highly chirped and pre-bunched beam leads to the stable generation of attosecond X-ray pulses. Furthermore, two-color attosecond pulses with sub-femtosecond separation can be produced by adjusting the energy distribution of the electron beam so that multiple FEL resonances occur at different locations within the beam. Such a tunable coherent attosecond X-ray sources may open up a new area of attosecond science enabled by X-ray attosecond pump/probe techniques.
\end{abstract}


\maketitle




Coherent attosecond pulses have significantly advanced modern science by allowing steering and tracking of electronic motion on an unprecedentedly fast time scale \cite{corkum2007attosecond, bucksbaum2007future, krausz2009attosecond, lepine2014attosecond, krausz2014attosecond, nisoli2017attosecond}. Many of these achievements are based on High-Harmonic Generation (HHG) sources \cite{agostini2004physics, sansone2011high, chini2014generation} that emit photons with energies tens of times that of the driving laser photon energy by generating high harmonics of an incident laser pulse of intensity $O(10^{15}~\watt/\centi\meter^2)$ when it interacts with rare gases. Recently even higher energy photons have been generated by another HHG mechanism by firing relativistic $\gtrsim10^{18}~\watt/\centi\meter^2$ laser pulses at a solid surface \cite{thaury2010high}. However, the conversion efficiency of the HHG photons tends to be low in the soft and hard X-ray regime, thereby limiting the range of applications of these HHG sources. On the other hand, X-ray free-electron lasers (XFELs) based on kilometer long radio-frequency (RF) accelerators \cite{saldin1999physics, huang2007review, pellegrini2016physics} can produce high power X-ray pulses by wiggling high-energy electrons inside a periodic magnetic undulator. Using this well proven approach to FELs recent work \cite{duris2020tunable} has demonstrated the generation of several hundreds of attosecond X-ray pulses with $\sim$100 GW peak power through sophisticated manipulation of several GeV electrons. 

The problem we wish to address here is, can the relativistic electron source and the undulator be miniaturized simultaneously to enable a truly milli-meter scale XFEL in the keV range? In this article we propose to collide a pre-bunched, highly chirped and ultra-high brightness nominally sub-100 MeV electron beam produced by an ultra-high gradient (GV/cm) plasma-based accelerator (PBA) \cite{PhysRevLett.43.267, chen1985acceleration, RevModPhys.81.1229, joshi2020perspectives} with an approximately 10 wavelengths long intense laser pulse that acts as an optical undulator. Only electrons with resonant energies can radiate coherently thereby emitting a short pulse. This self-selection process can tolerate large jitters of the electron beam (energy, pointing and position) produced from PBA, which hinders the operation of current PBA driven XFELs \cite{wang2021free, pompili2022free}.

\begin{figure*}
\includegraphics[width=0.8\textwidth]{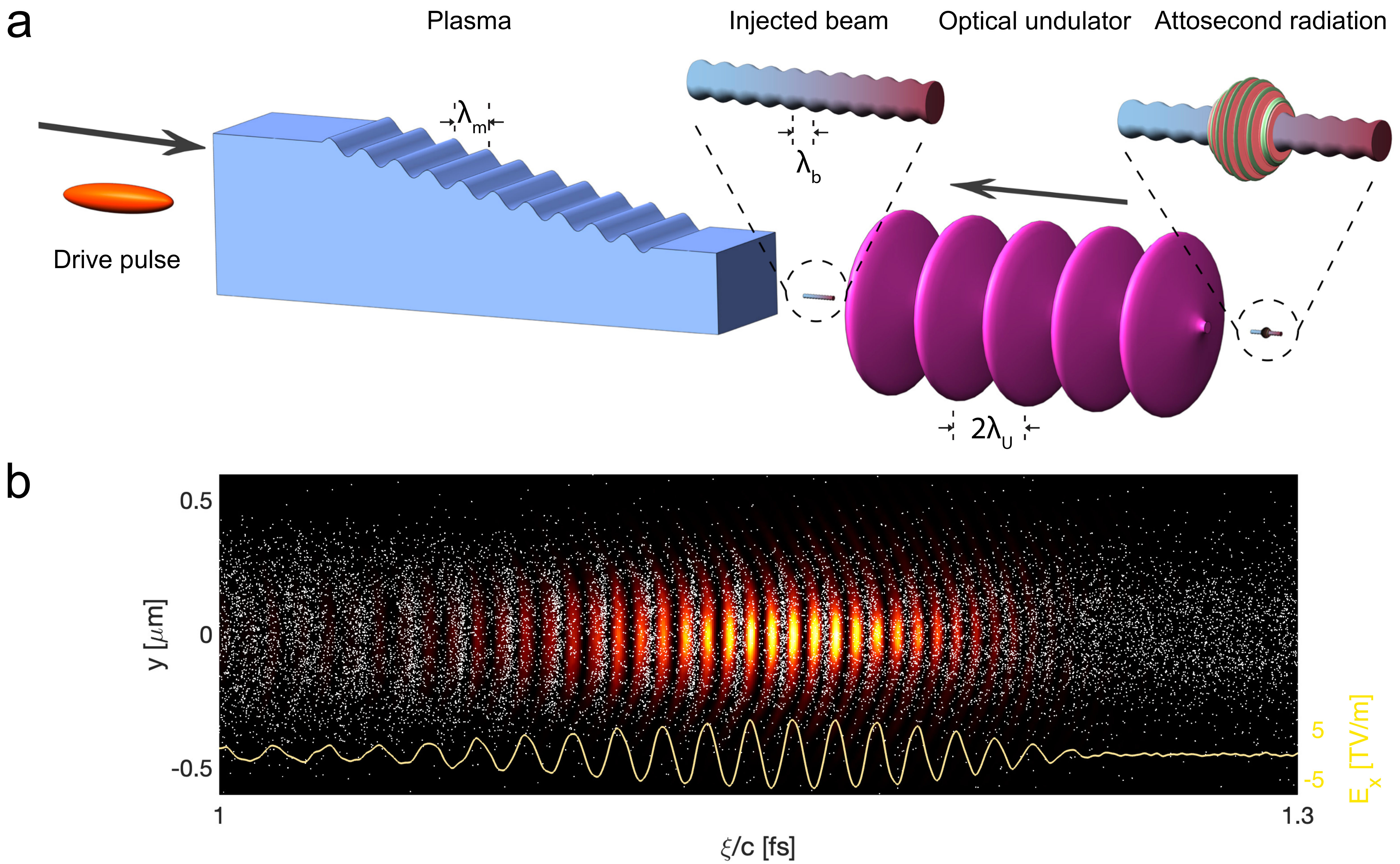}
\caption{\label{fig: concept} \textbf{Illustration of attosecond pulse generation. a,} A highly relativistic electron beam driver propagates through a density modulated plasma downramp to inject a pre-bunched, highly chirped and ultra-bright electron beam which collides with a laser pulse at the exit of the plasma to generate an attosecond pulse. \textbf{b,} 3D PIC simulation results: Beam energy 1 GeV, peak current 34 kA, upper/lower shelf plasma density $2.17/1.97\times 10^{19}~\centi\meter^{-3}$, ramp length 60 $\micro\meter$ and plasma density modulation period 400 nm. Undulator is a 10 $\micro\meter$ wavelength laser pulse. The electrons bunching (white dots) on X-ray radiation scale and the radiation intensity $E_x^2$  (black-red-yellow) at $x=0$ plane when it saturates ($z=10\lambda_\mathrm{U}$). The yellow line represents the electric field $E_x$ along $y=0$. }
\end{figure*}

Electrons generated from PBA are characterized by unique parameters that are not accessible for RF accelerators, e.g., ultra-high brightness \cite{xu2017high, dalichaouch2020generating, li2022ultrabright} and large energy chirps, which may open previously inaccessible regimes for XFELs. The generation of ultraviolet radiation from the exponential gain from an electron beam produced in PBA propagating in a magnetic undulator has been recently demonstrated \cite{wang2021free, pompili2022free}. Attosecond radiation produced by compressing beams from PBA using a magnetic chicane and then radiating in a magnetic undulator has been proposed \cite{emma2020terawatt}. 

In this article, we propose a radically different ultra-compact XFEL concept, depicted in Fig. 1, that is based on an unexplored regime enabled by PBA and an optical undulator. There are several key features that distinguish this concept from other conventional or PBA based XFEL concepts. First, we propose pre-bunching a highly chirped electron beam on an X-ray wavelength scale in the PBA. Due to the huge acceleration gradients in PBA, the injected beam can be characterized by $10s$ of MeV/fs energy chirp, i.e., the electron’s energy strongly depends on its axial phase within the plasma wake. When a pre-bunched beam with a large energy chirp oscillates inside an undulator, the electrons with resonant energy at the bunching wavelength radiate coherently \cite{gover2019superradiant} while electrons at other energies begin emitting photons at a much lower fluctuation level. As a result, only a small portion of the beam radiates a fully coherent attosecond pulse with a stable energy and spectrum, which is in contrast to the ultrafast pulses generated from RF accelerator based XFELs \cite{duris2020tunable, huang2017generating, zhang2020experimental} in the self-amplified spontaneous emission mode whose pulse energy and spectrum fluctuate significantly. Second, we also propose the use of a powerful, sub-picosecond class laser pulse as an optical undulator \cite{elias1979high, gover1982feasibility}. An undulator wavelength of a few microns reduces the resonant energy of the electron beam needed to produce X-rays from a few GeV to sub-100 MeV, and the direct collision of the electron beam with the undulator laser pulse eliminates the need to have a complex electron beam transport line (see Fig 1) to match the beam from the plasma to the magnetic undulator \cite{wang2021free, pompili2022free} without significant degradation of the beam quality \cite{PhysRevSTAB.17.054402, xu2016matching}. Finally, the ultra-high brightness beam allows the FEL instability to grow rapidly within an undulator wavelength and the radiation saturates before the beam diffracts significantly. The distance over which the radiation power grows by a factor of $e$ (aka gain length) is comparable to the undulator wavelength, which is orders of magnitude shorter than possible in conventional XFELs \cite{saldin1999physics, huang2007review, pellegrini2016physics, duris2020tunable, huang2017generating, zhang2020experimental}. The self-selection mechanism of this novel FEL instability can tolerate $10s$ of MeV energy jitter of the beams while the use of an optical undulator without transverse focusing can accept beams with $10s$ of degree pointing jitter and $10s$ of micron transverse position jitter. In PBA, there can be large jitters to the electron beam characteristics, thus these tolerances are critical for stable operation of the radiation source.

Additionally, attosecond harmonic pulses can be generated with sub-femtosecond temporal separation by different parts of the chirped beam if its energy range covers the resonant energies of harmonics. These synchronized pulses can be used as multiple probes with different colors or enable attosecond pump/probe experiments, thus greatly broadening the application range of the proposed concept. By utilizing ultrashort electron bunches produced from laser plasma accelerators \cite{tooley2017towards} as drivers and laser pulses with micron wavelength as undulator, this scheme can be scaled to hard X-ray regime to produce several attosecond pulses. 
\\
\\

\noindent
\textbf{\Large Results}

\noindent
To demonstrate the new physics of the proposed PBA based XFEL, we present high fidelity simulation results on the generation of $10s$ of attosecond X-ray pulses with several GW peak power. The simulations model each of the three components of the concept described above including the collision of highly chirped, pre-bunched and ultra-bright electron beam, self-consistently produced in a PBA stage, with an optical undulator. The three-dimensional (3D) particle-in-cell (PIC) code OSIRIS \cite{fonseca2002high} is used to simulate the entire concept, including a new regime of the FEL instability which cannot be accurately modeled with standard FEL codes, such as Genesis 1.3 \cite{reiche1999genesis}. The concept utilizes an extremely bright beam, leading to a very large FEL amplification bandwidth, and this together with the large energy chirp and strong space charge effects, leads to an XFEL regime that cannot be described with the standard FEL theory or let alone modeled with standard simulation codes.

\begin{figure*}
\includegraphics[width=0.8\textwidth]{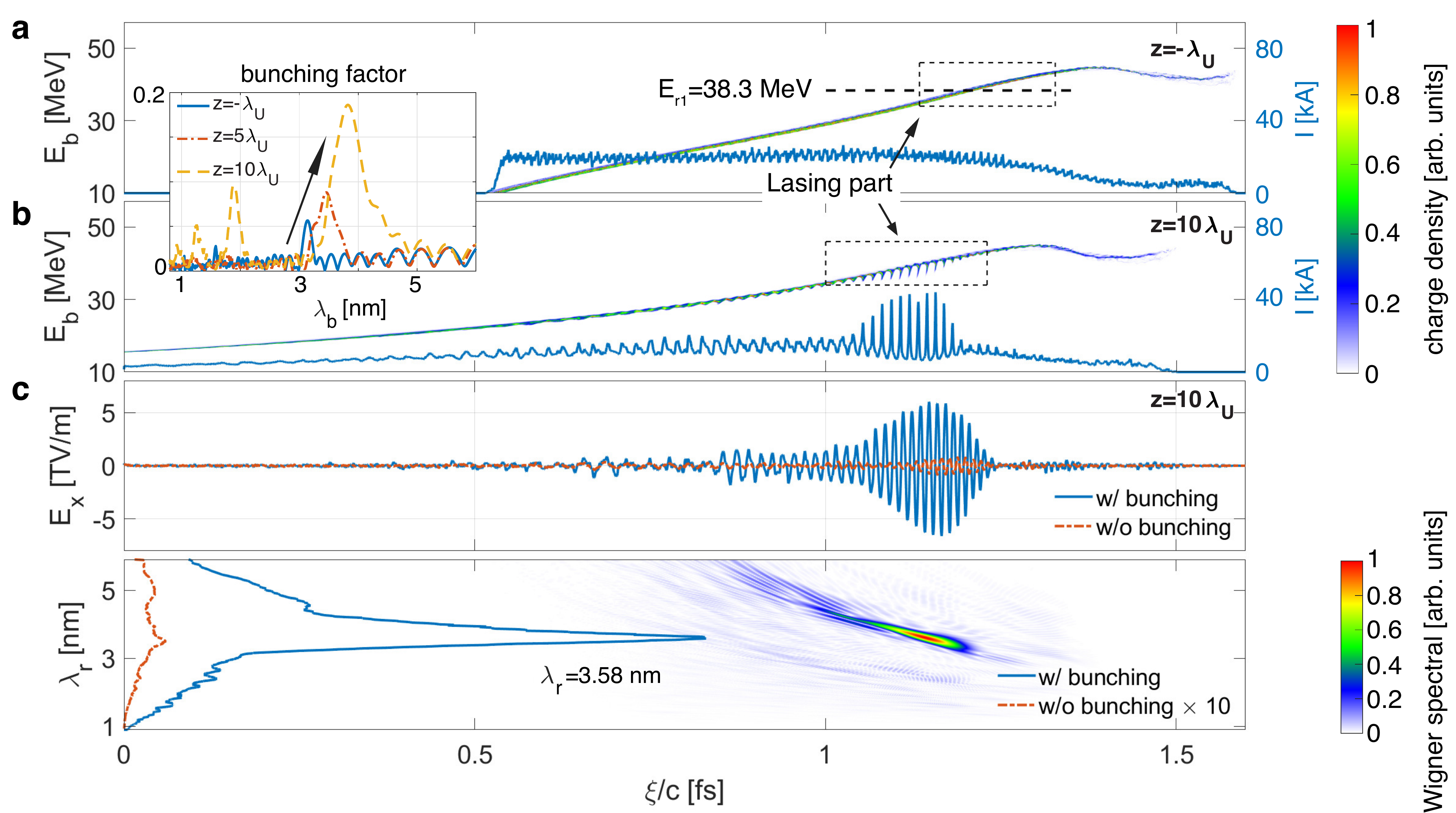}
\caption{\label{fig: as pulse} \textbf{The properties of the beam and the attosecond pulse. a, b,} The longitudinal phase space of the beam and its current profile (blue) when it exits the plasma (\textbf{a}) and when the radiation is saturated (\textbf{b}). The bunching factor at three different positions inside the undulator is shown in the inset where the growth and shift of the bunching factor can be seen. \textbf{c,} The on-axis electric field and its Wigner function when the radiation is saturated. The results at the same location when the pre-bunching is absent are also shown as a comparison. Note the beam moves from left to right.}
\end{figure*}

\noindent
\textbf{Bunched beam generation in PBA.} Recently a plasma downramp along the driver’s propagation direction ($z$) has been proposed to decrease the phase velocity of the wake from the driver’s velocity sufficiently as to trap plasma electrons in a nonlinear wake \cite{PhysRevA.33.2056, PhysRevLett.86.1011, PhysRevE.58.R5257}. Numerical simulations have shown that beams with ultra-high brightness \cite{xu2017high} can be injected into 3D nonlinear blowout wakes \cite{PhysRevLett.96.165002}. If a small amplitude sinusoidal density modulation is superimposed on the downramp, the wake expands and shrinks periodically thereby turning the electron injection on/off during the expansion/contraction while the downramp maps the discrete injection to different axial phases leading to a bunched current profile with a period that is significantly compressed compared to that of the density modulation \cite{xu2022generation}.

Such a plasma density modulation can be created by using the ponderomotive force from the interference of two linearly polarized (LP) degenerate laser pulses \cite{xu2022generation, sheng2003plasma, zhang2021ionization}. A premodulated plasma downramp is used in the following simulations to save computational cost. In the illustration shown in Fig. \ref{fig: concept}a, an electron beam driver excites a wake. As the driver traverses the density-modulated ramp, electrons are injected periodically to form a current modulated (bunched) beam \cite{xu2022generation}. Note that the drive-electron beam itself needs not be ultrabright to initiate the FEL instability. In fact, drive beams that are suitable to excite wakes can be produced using either a laser plasma accelerator \cite{kurz2021demonstration} or an RF accelerator. It is during this bunching process arising from periodic injection in the wake excited in the density modulated downramp that the plasma accelerator acts as a brightness transformer generating a string of high current electron micro-bunches on X-ray wavelength scale that can radiate superradiantly when they encounter an intense counter-propagating laser pulse. The compression factor which is defined as the ratio of the plasma density modulation period and the bunching period of the injected beam can reach several hundred depending on the density gradient of the ramp. For the parameters used here (see Methods), the injected beam is bunched at $\lambda_\mathrm{b}\approx 3.2$ nm with a compression factor of 125 as shown by its current profile in Fig. \ref{fig: as pulse}a. Due to the mapping between the electrons initial axial position ($z_\mathrm{i}$) and their final position (phase) inside the wake ($\xi\equiv ct-z$), electrons injected earlier are accelerated a longer distance and reside at the head of the beam, thus there is a 30 MeV/fs linear energy chirp \cite{xu2017high} along the beam at the end of the ramp as shown in Fig. \ref{fig: as pulse}a.

\noindent
\textbf{Attosecond pulse generation in an optical undulator.} As an example of an optical undulator we here employ an appropriately delayed LP CO$_2$ laser pulse with wavelength $\lambda_\mathrm{CO_2} = 10~\micro\meter$ and normalized vector potential $a_\mathrm{CO_2}= \frac{eE_\mathrm{CO_2}\lambda_\mathrm{CO_2}}{2\pi mc^2} = 3.52$ collides with the bunched beam as it exits the downramp, where $E_{\mathrm{CO_2}}$ is the laser's electric field. Such a laser pulse is equivalent to a magnet undulator with $\lambda_\mathrm{U}=5~\micro\meter$ and $K=3.52$, where $K=\frac{eB_\mathrm{U}\lambda_\mathrm{U}}{2\pi mc}$ is the undulator normalized vector potential amplitude and $B_\mathrm{U}$ is the magnetic field on axis. The electrons with energy $\gamma_\mathrm{b} mc^2$ oscillate along the laser polarization direction ($x$) with an amplitude $\frac{K\lambda_\mathrm{U}}{2\pi \gamma_\mathrm{b}}$ and radiate. The resonance condition gives the radiation wavelength as \cite{saldin1999physics, huang2007review, pellegrini2016physics}
\begin{align}
\lambda_\mathrm{r} = \frac{\lambda_\mathrm{U}}{2q\gamma_\mathrm{b}^2}\left( 1 + \frac{K^2}{2}\right)
\end{align}
where $q$ is the harmonic number. For a pre-bunched beam with wavelength $\lambda_\mathrm{b}=3.2$ nm and $q=1$, electrons with energy $\gamma_\mathrm{r1}\approx 75$ (37.5 MeV) radiate superradiantly while others radiate incoherently. Thus, these resonant electrons can develop the FEL instability much faster. This is clearly seen in Figs. \ref{fig: as pulse}a and b where the electron beam’s $(p_z, \frac{\xi}{c})$ phase space and axial current profile are shown as the beam enters the laser undulator and after saturation. In Fig. \ref{fig: as pulse}b, the large current modulation and corresponding loss of beam energy are clearly seen only for the part of the beam near the resonant energy. The distribution of the electrons and the radiation intensity when it saturates are shown in Fig. \ref{fig: concept} where most electrons have slipped to the positions where the radiation intensity is close to zero. 

The bunching factor of the lasing part of the beam, $b=|\sum_{j=1}^{N_\mathrm{b}}\mathrm{exp}(ikz_j)|/N_\mathrm{b}$ grows from an initial value of 0.06 to 0.19. Here $N_\mathrm{b}\approx 2.3\times 10^7$ is the total number of electrons in the region encompassed by the dashed rectangle in Figs. \ref{fig: as pulse}a and b. The on-axis electric field of the radiation and its Wigner function are shown in Fig. \ref{fig: as pulse}c. It can be seen that a 3.58 nm radiation pulse with 7.1 TV/m peak field, 7.6 GW peak power and 96.2 as full width at half maximum (FWHM) duration (corresponding to $0.64\times 10^{10}~0.35 \kilo\electronvolt$ photons) is generated. The radiated pulse has a $-5.34$ nm/fs chirp which is caused by the chirp of the beam, i.e., $\frac{\Delta \lambda_\mathrm{r}}{\lambda_\mathrm{r}} \approx -2 \frac{\Delta \gamma_\mathrm{b}}{\gamma_\mathrm{b}}$. The FWHM spectral width is 66 eV which can support a Fourier transform limited pulse with 27.4 as duration. For comparison, the radiation pulse and its spectra when the pre-bunching is absent are also shown in Fig. \ref{fig: as pulse}c (red). Without pre-bunching, different parts of the beam grow from noise independently; thus, the field is weak, and the spectrum is broad \cite{krinsky2003frequency}. 

\begin{figure*}
\includegraphics[width=0.7\textwidth]{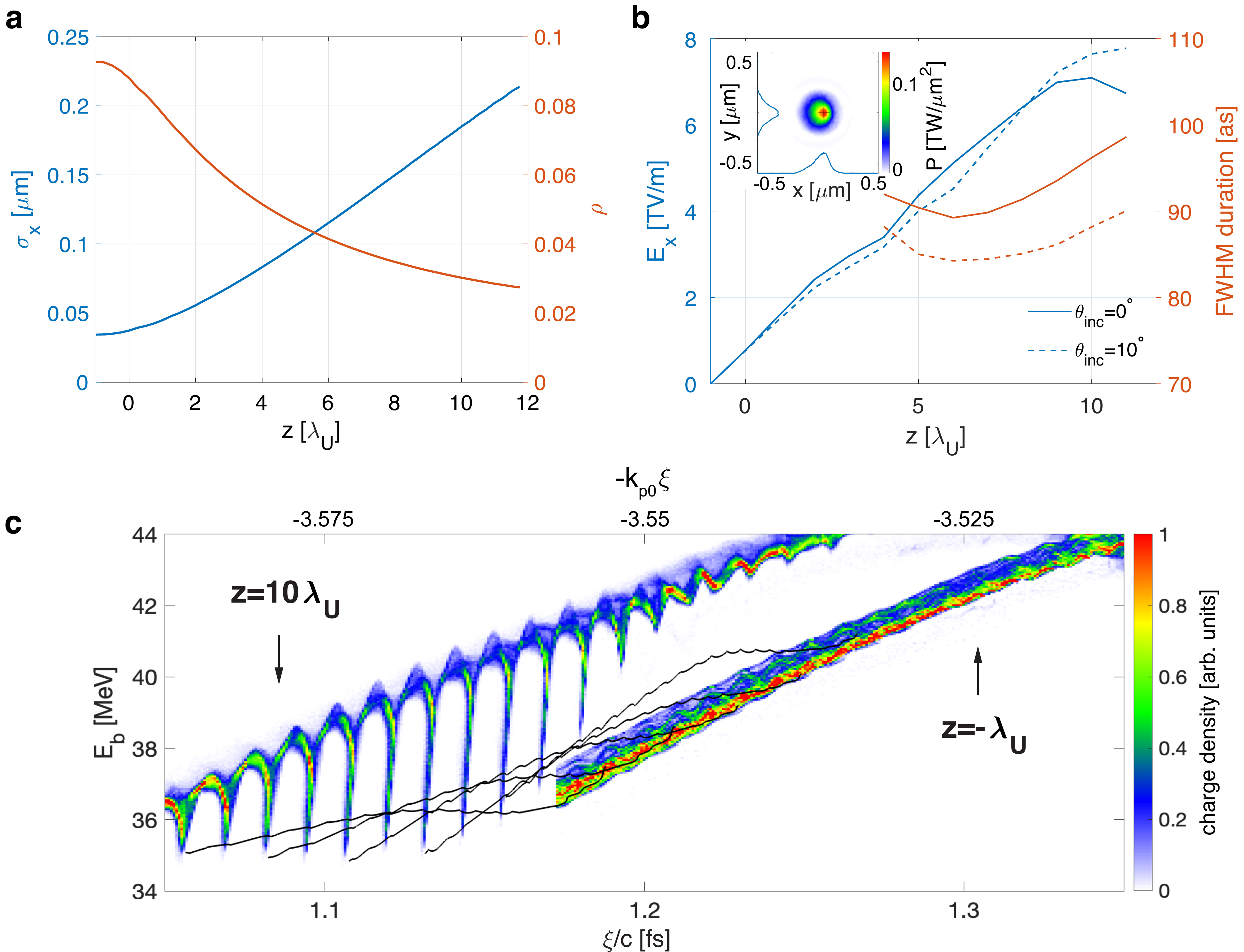}
\caption{\label{fig: dynamics} \textbf{Attosecond pulse generation inside an optical undulator. a,} The evolution of the beam spot size (blue) and the growth rate (red) of the instability. \textbf{b,} The evolution of radiation field amplitude and its duration. The inset shows the power density distribution at its peak ($t=1.172~\femto\second,z=10\lambda_\mathrm{U}$) and the lineouts. A case where the CO2 laser collides with the beam with 10° angle is shown to demonstrate the robustness of this scheme. Note that the intensity of the laser is decreased ($a_\mathrm{U}=3.51$) to compensate the wavelength change due to the angle. \textbf{c,} The energy loss and slippage of the lasing electrons where the black lines show the trajectories of electrons which lose a significant amount of energy.}
\end{figure*}

Due to the ultra-high focusing gradient inside the plasma wake, the output beam is tightly focused to a 34 $\nano\meter$ spot size with a peak density of $n_\mathrm{b}\approx 5.9\times10^{22}~\centi\meter^{-3}$. Such a high-density beam can drive the FEL instability with a normalized growth rate defined by the Pierce parameter \cite{saldin1999physics, huang2007review, pellegrini2016physics} which for the parameters simulated can be as large as $\rho= \left( \frac{e^2K^2[\mathrm{JJ}]^2n_\mathrm{b}}{32\epsilon_0\gamma_\mathrm{r1}^3mc^2k_\mathrm{U}^2} \right)^{1/3} \approx 0.09$, which is orders of magnetic higher than that in RF accelerator based XFELs ($\sim0.001$). The corresponding gain length is $L_\mathrm{g} = \frac{\lambda_\mathrm{U}}{4\pi\sqrt{3} \rho} \approx 2.6~\micro\meter$ which is much less than the undulator period. Here $\epsilon_0$ is the vacuum permittivity and $[\mathrm{JJ}]=0.772$ is the coupling factor between the electron beam and the field \cite{saldin1999physics, huang2007review, pellegrini2016physics}. Due to its high density and low energy, the plasma oscillations between the current peaks of the beam occur on a length which is comparable to the gain length ($k_\mathrm{pb}^{-1}\approx 4.9~\micro\meter$) and thus contributes to the instability. The FEL instability is in the transition between the Compton ($k_\mathrm{pb}^{-1}\ll L_\mathrm{g}$) and the Raman regime ($k_\mathrm{pb}^{-1} \gg L_\mathrm{g}$)  \cite{gover1981unified}.

Only electrons with energies corresponding with resonant frequencies inside the gain bandwidth can contribute to the radiation, thus the duration of the radiation pulse can be estimated as $\sigma_\tau \sim \frac{\rho}{(\mathrm{d}\gamma_\mathrm{b}/\mathrm{d}t)/\gamma_\mathrm{b}} $. The spectra as a function of propagation distance found in the inset of Fig. \ref{fig: as pulse} shows the bunching wavelength of the lasing electrons moves from initially 3.15 nm to 3.84 nm at saturation. This shift is mainly ascribed to the microbunching decompression of a chirped beam in an undulator since the high energy electrons move faster than low energy electrons (see Supplementary Material). 

As the beam exits the plasma and expands inside the laser pulse, its spot size increases and thus the density decreases. This leads to a decrease in the the Pierce parameter from $\rho\approx$0.09 to 0.03 in 10 undulator periods as shown in Fig. \ref{fig: dynamics}a while the FEL instability grows to saturation during this time as shown in Fig. \ref{fig: dynamics}b. The longitudinal phase space ($p_z, \frac{\xi}{c}$) of the lasing part of the beam is shown for  $z=-\lambda_\mathrm{U}$ and 10$\lambda_\mathrm{U}$ in Fig. \ref{fig: dynamics}c. The electrons that lose energies are initially separated by a radiation wavelength and each electron loses as much as $11\%$ of their initial energy (4.5 MeV), which corresponds to $1.3\times 10^{4}~0.35~\kilo\electronvolt$ photons per electron. Therefore even though the number of electrons per microbunch is small [$O(100~\femto\coulomb)$] the total number of photons per pulse can be $\sim10^{10}$. The trajectory of several selected electrons is also shown illustrating that the rate of energy loss varies across the beam and in time. As can be seen, the position of the beam at $z=-\lambda_\mathrm{U}$ and $10~\lambda_\mathrm{U}$ has slipped due to the well-known phase slippage in the FEL instability. Fig. \ref{fig: dynamics}b shows the evolution of the radiation field and its duration. The field grows roughly linearly and not exponentially since the transverse expansion of the beam leads to a continuously decreasing growth rate (see Supplementary Material) and then saturates. The pulse duration starts to increases after 6 periods because the slippage. The transverse distribution of the radiation power density is shown in the inset with $\sigma_{x,y} \approx 80~\nano\meter$. The absence of the transverse focusing force along the optical undulator allows the beam to be incident with a large angle $\theta_\mathrm{inc}$ which only leads to a decrease of the undulator period as $\lambda_\mathrm{U}\frac{1+\mathrm{cos}\theta_\mathrm{inc}}{2}$. An example with $\theta_\mathrm{inc}=10\degree$ is shown in Fig. \ref{fig: dynamics}b where a similar radiation pulse is produced.

\noindent
\textbf{Generation of two attosecond pulses with sub-femtosecond separation}. In the above results, the maximum energy of the beam is close to the resonant energy $E_\mathrm{r1}$ of the fundamental bunching wavelength and an isolated attosecond pulse with the bunching wavelength is generated. If the beam is accelerated further inside the plasma, it can cover energies that are not only resonant with the fundamental but also resonant with harmonics of the pre-bunched wavelength. Harmonics can be seeded if the bunching is sufficiently strong such that the bunching factor contains harmonics. Depending on the range of energies of the beam different regions of the beam will be resonant with different harmonics and radiate independently to generate several fully coherent attosecond pulses with multiples of the fundamental wavelength and controllable separations on an sub-femtosecond scale. 

\begin{figure*}
\includegraphics[width=0.8\textwidth]{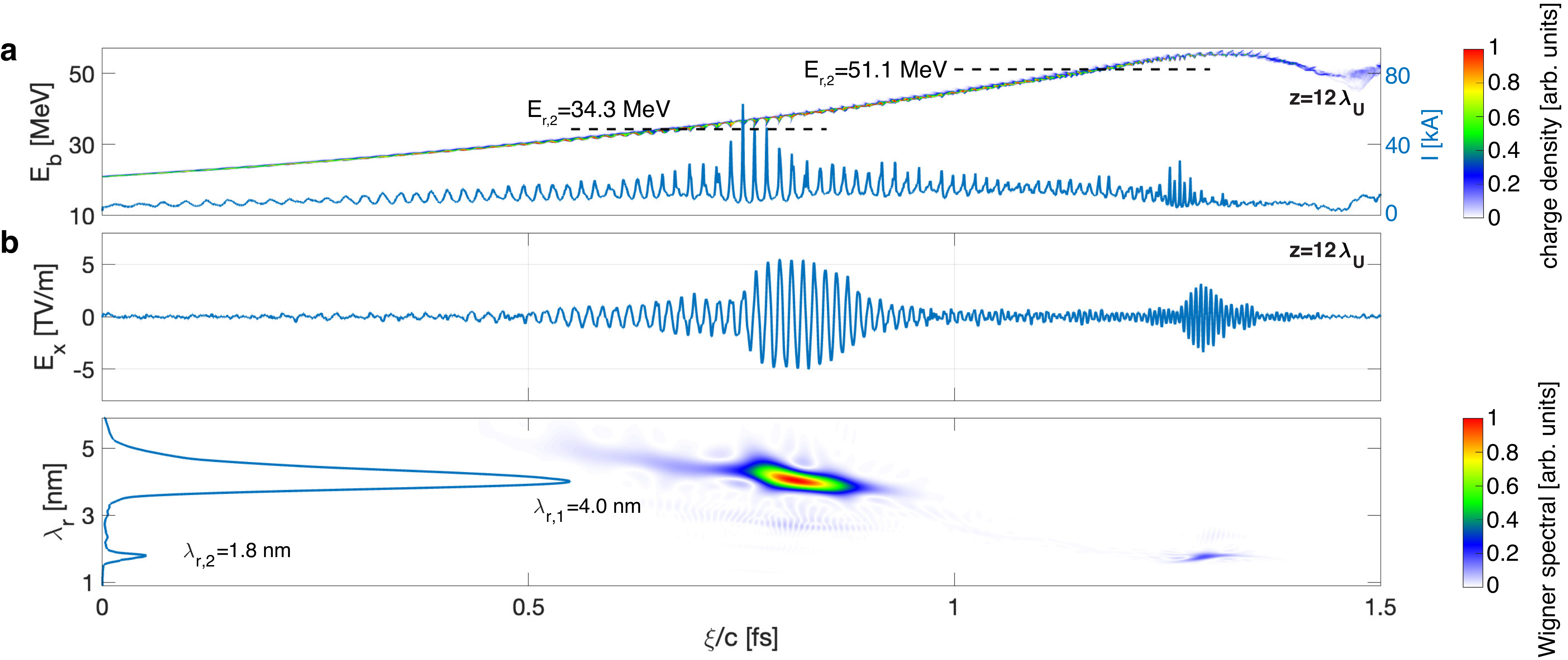}
\caption{\label{fig: 2 pulses} \textbf{Generation of three attosecond pulses with different wavelengths. a,} The longitudinal phase space at $z=12\lambda_\mathrm{U}$ and the current profile of the beam. The resonant parts of the three radiation pulses are shown by black dashed lines. \textbf{b,} The on-axis electric field and its Wigner function. A CO$_2$ laser pulse with $\lambda_\mathrm{CO_2}=10~\micro\meter, a_\mathrm{CO_2}=3.52$ is used and the beam is accelerated with $12~\micro\meter$ after the ramp to reach a high energy. }
\end{figure*}

Fig. \ref{fig: 2 pulses} shows an example where two attosecond pulses are generated. The 37.5 MeV beam is further accelerated in a density plateau of a length of 12 $\micro\meter$ that follows the ramp to a maximum energy of 57 MeV which not only covers the fundamental but also the 2nd ($E_\mathrm{r2}$=51.1 MeV) harmonics. The longitudinal phase space of the beam after 10 undulator periods is shown in Fig. \ref{fig: 2 pulses}a where two regions lose energy. As shown in Fig. \ref{fig: 2 pulses}b, two attosecond pulses (93 as and 30 as) with different wavelengths (4.0 nm and 1.8 nm) separated by 477 as are produced. The wavelengths are not exact harmonics because the shift of the fundamental wavelength during the propagation is more than that of the 2nd harmonic (see Supplementary Material). The peak powers of these two pulses are 6.6 GW and 1.5 GW and they contain $0.65\times10^{10}$ 0.31 keV photons and $0.19\times10^9$ 0.69 keV photons, respectively. Such a sequence of two-colored pulses can enable attosecond pump and attosecond probe studies or probe an ultra-fast process to obtain a “movie” with attosecond resolution in a single shot.
\\
\\

\noindent
\textbf{\Large Discussion}

\noindent
The properties of coherent, attosecond X-ray pulses generated from the interaction of a highly chirped and pre-bunched beam with an optical undulator can be tuned easily. The duration of each pulse and their separation are inversely proportional to the local chirp of the beam which can be controlled by the gradient of the ramp \cite{xu2017high, xu2022generation}. An example, where a beam with a larger energy chirp generates 38 as and 15 as pulses with 66 as separation is shown in the supplement. By controlling the acceleration distance inside the plasma, the energy distribution of the beam can be varied which determines the number of the generated pulses. Extensive simulation scans have found that the current of the self-injected beam produced from downramp injection is around half of the drive beam current \cite{xu2017high, dalichaouch2020generating, li2022ultrabright}, thus the peak power of the radiation can be increased if a drive beam with higher current is used. The wavelength of the radiation which is approximately equal to the bunching wavelength of the electron beam can be continuously tuned by adjusting the ramp gradient and/or the plasma density modulation period \cite{xu2022generation}. 

A short and tightly focused GeV-class electron beam driver is used in the above simulations to effectively excite a nonlinear plasma wake and produce pre-bunched beams. Adequate electron beams might exist in the future at both FACET-II \cite{Yakimenko2019PRAB} and FlashForward \cite{FLASHForward2019}. However, such a short electron beam driver could be produced in a separate laser plasma accelerator stage. Alternatively, intense laser pulses could replace the electron beam drivers and be used directly to produced pre-bunched beams which lead to simpler design of the proposed concept \cite{xu2022generation}. Either an electron beam driver from a laser plasma accelerator or a laser pulse driver may make proposed concept accessible to experimental verification because there are almost a 100, hundred TW class laser systems in the world.

The above radiation generation scheme will lead to a more compact XFEL compared to other PBA-based schemes \cite{wang2021free, pompili2022free, emma2020terawatt} that nevertheless use a conventional magnetic undulator and multiple transport components. The proposed scheme is free of the need to transport and match the beam to the undulator since the beam interacts with the optical undulator immediately after the plasma. It can tolerate large transverse angular and position jitters since the undulator is only tens of optical wavelengths long and has a much larger spot size than the electron beam. Since this process automatically selects the lasing portion from the whole beam, it can tolerate a beam with a large energy spread or significant shot-to-shot energy jitter as long as the resonant energy is covered by the beam. The gain bandwidth of the FEL instability is roughly equal to the Pierce paramete \cite{saldin1999physics, huang2007review, pellegrini2016physics} which is $\rho\approx 0.09$ for the proposed concept. Such a large bandwidth can tolerate a large fluctuation of the optical undulator intensity. These advantages make this ultra-compact XFEL scheme particularly attractive. 
\\
\\

\noindent
\textbf{\Large Methods}

\noindent
\textbf{Particle-in-cell (PIC) simulations.} The simulations are carried out using the fully relativistic, electromagnetic particle-in-cell (PIC) code OSIRIS \cite{fonseca2002high}. The injection of the bunched beam in a density modulated ramp is modeled in cylindrical geometry. A simulation window moving with speed of light has dimensions of $13.5\times14.4~\micro\meter$ with $18000\times 6144$ cells in the $z$ and $r$ directions, respectively. This corresponds to cell sizes of $\Delta z=0.75~\nano\meter$ and $\Delta r=2.34~\nano\meter$. The time step is $\Delta t=\frac{\Delta z}{2c}=1.25~\atto\second$. There are 8 macro-particles per cell to represent the plasma electrons and beam electrons. A Maxwell solver with an extended stencil \cite{li2017controlling} (16 coefficients) is used to model the bunched beam generation with high-fidelity \cite{xu2020numerical}. The 1 GeV electron beam driver has a tri-gaussian distribution with a spot size of $0.6~\micro\meter~(0.5 k_\mathrm{p0}^{-1})$, a duration of $0.84~\micro\meter~(0.7~k_\mathrm{p0}^{-1})$ and a 34 kA peak current. The plasma starts with a $30~\micro\meter~(25~k_\mathrm{p0}^{-1})$ sinusoidal upramp where the density increases from zero to $1.1n_\mathrm{p0}~(n_\mathrm{p0}=1.97\times10^{19}~\centi\meter^{-3})$, then a $30~\micro\meter~(25~k_\mathrm{p0}^{-1})$ long plateau, then a $60~\micro\meter~(50~k_\mathrm{p0}^{-1})$ long downramp where the density is decreased linearly from $1.1n_\mathrm{p0}$ to $n_\mathrm{p0}$. A sinusoidal density modulation with 400 nm period and $0.002n_\mathrm{p0}$ amplitude is superimposed along the ramp. An acceleration stage of length $12~\micro\meter~(10~k_\mathrm{p0}^{-1})$ is used in Fig. \ref{fig: 2 pulses} to improve the electrons’ energy while it is absent in Figs. \ref{fig: as pulse} and \ref{fig: dynamics}. Results from a simulation conducted with a finer grid size ($\Delta z=0.3~\nano\meter, \Delta r=0.59~\nano\meter)$ to resolve the 2nd harmonic are shown in Fig. \ref{fig: 2 pulses}. 

The distributions of the bunched beams generated from the $r-z$ simulations described above are interpolated onto 6D phase space by giving each particle a random phase in $x-y$ and $p_x-p_y$ space, respectively. Then these 6D phase space distributions are imported to full-3D PIC simulations where they interact with an optical undulator to generate attosecond pulses. For the case without bunching shown in Fig. \ref{fig: as pulse}c, a random axial offset with 4 nm amplitude (slighter longer than the bunching wavelength) is added to the electrons; axial positions to eliminate the bunching. Since the beam energy varies significantly across the beam (from $\sim10$ MeV to $\sim60$ MeV) and the growth rate of the instability is so large that the radiation grows significantly even in one undulator period, the generation of the radiation cannot be modeled using current major FEL codes, like GENESIS 1.3 \cite{reiche1999genesis}. We use PIC code OSIRIS to model this process in full 3D geometry. Such simulations require sufficient resolution to accurately model the bunching and X-ray wavelength. A simulation window moving with speed of light has dimensions of $0.48\times1.2\times1.2~\micro\meter^3$ with $3200\times320\times320$ cells in the $z,x$ and $y$ directions, respectively. This corresponds to cell sizes of $\Delta z=0.15~\nano\meter$ and $\Delta x=\Delta y=3.75~\nano\meter$. The time step is $\Delta t=\frac{ \Delta z}{2c}=0.25~\atto\second$. A customized Maxwell solver with an extended stencil \cite{li2017controlling} (16 coefficients) is used to model the attosecond pulse generation with high-fidelity \cite{xu2020numerical}. The optical undulator is described by a prescribed 1D field profile which is reasonable since its spot size is much larger than the transverse size of the simulation box. It starts with a profile as $a_\mathrm{CO_2} (k_\mathrm{CO_2} z - \omega_\mathrm{CO_2} t)\mathrm{exp}\left[ - \frac{ (k_\mathrm{CO_2} z - \omega_\mathrm{CO_2} t)^2 }{2} \right] $ and then a constant envelope.

\noindent
\textbf{Data Analysis.} The Wigner distribution of the radiation electric field is obtained by using the ‘wvd’ function provided by MATLAB. The one shown in Fig. \ref{fig: 2 pulses}c use a Kaiser window with shape factor $\beta=20$ and width 500.3 as (1001 data points) for the time domain and a Kaiser window with shape factor $\beta=20$ and width 313 PHz (1001 data points) for the frequency domain to avoid the interference between the two attosecond pulses with different wavelengths.

\bibliography{refs_xinlu}

\noindent
\textbf{Acknowledgements}
This work was supported by the National Natural Science Foundation of China (NSFC) (No. 11921006) and the National Grand Instrument Project (No. 2019YFF01014400), the U.S. Department of Energy under Contracts No. DE-AC02-76SF00515 and No. DE-SC0010064, the U.S. National Science Foundation under Grants No. 2108970, and the DOE Scientific Discovery through Advanced Computing (SciDAC) program through a Fermi National Accelerator Laboratory (FNAL) subcontract No. 644405. The simulations were performed on the resources of the National Energy Research Scientific Computing Center (NERSC), a U.S. Department of Energy Office of Science User Facility located at Lawrence Berkeley National Laboratory.

\bigskip
\noindent
\textbf{Author Contributions}

\noindent

\bigskip
\noindent
\textbf{Competing interests}

\noindent
The authors declare no competing financial interests.

\bigskip
\noindent
\textbf{Additional information}

\noindent

\end{document}